\documentclass[aps,amsmath,amssymb,nofootinbib]{revtex4}
\usepackage{graphicx,graphics,epsfig,subfigure}

\newcommand{\bi}{\begin{itemize}}
\newcommand{\ei}{\end{itemize}}
\newcommand{\be}{\begin{eqnarray}}
\newcommand{\ee}{\end{eqnarray}}
\newcommand{\ba}{\begin{array}}
\newcommand{\ea}{\end{array}}
\newcommand{\bc}{\begin{center}}
\newcommand{\ec}{\end{center}}
\newcommand{\bt}{\begin{tabular}}
\newcommand{\btab}{\begin{table}}
\newcommand{\et}{\end{tabular}}
\newcommand{\ds}{\displaystyle}
\begin{document}

\title{Long-Wavelength Modes of Cosmological Scalar Fields}

\author{Marcin Jankiewicz}
\email{m.jankiewicz@vanderbilt.edu}
\affiliation{Department of Physics and Astronomy, Vanderbilt
University, Nashville, TN 37235, USA}

\author{Thomas W. Kephart} 
\email{thomas.w.kephart@vanderbilt.edu}
\affiliation{Department of Physics and Astronomy, Vanderbilt
University, Nashville, TN 37235, USA}

 \date{\today}

\begin{abstract}
We give a numerical analysis of long-wavelength modes in the WKB approximation of cosmological scalar fields coupled to gravity via $\xi\phi^{2} R$. Massless fields are coupled conformally at $\xi=1/6$. Conformality can be preserved for fields of nonzero mass by shifting $\xi$. We discuss implications for density perturbations.
\end{abstract}

\maketitle

\section {Introduction}
Scalar fields have played a major role in attempts to model the early Universe. In particular, nearly every incarnation of the inflation scenario has relied on scalars to generate vacuum energy and in turn exponential expansion and density fluctuations. Many of these models rely on slow roll potentials, i.e., potentials that are nearly flat where the scalar masses can be very small.

Recently, horizon size and super horizon size density perturbations have been studied intensively, because of their importance for understanding low $\ell$ modes (see \cite{Kolb:2005me} and references therein) in the WMAP data \cite{Bennett:2003bz}.
Long wavelength scalar field modes have interesting properties when the wavelength is on the order of the horizon size $c H_{0}^{-1}$. One finds dispersion and diffraction effects that depend on the scalar mass and it's  coupling to gravity. The generic lagrangian for a scalar in a Friedmann-Robertson-Walker (FRW) Universe is 
\be {\cal L} =  g^{\mu\nu} \partial_{\mu} \phi\partial_{\nu}\phi +\xi\phi^{2} R -V(\phi)\,.\ee
If $V(\phi)$ contains no dimensionful parameters, then the scalar field is conformally coupled when $\xi=\frac{1}{6}$. Conformal invariance can be broken by including a mass term in $V(\phi)$.
 Here we assume the local (Minkowski limit) real $\phi^{4}$ theory is renormalizable. While this is not completely general, it is sufficient for our purposes.
One could easily generalize our analysis to complex fields or fields in irreducible representations of some continuous symmetry group. 

One expects the scalars to be 
 an integral part of any realistic model, so for instance, if the overarching theory is based on strings with a local or global SUSY preserved down to some scale, then the scalars will be components of some superfield $\Phi$ contributing to the superpotential $W(\Phi)$. This will put constraints on $V(\phi)$. In particular, flat directions could result (regions of moduli space where the scalar mass vanishes) and lead to massless or nearly massless modes, where for example SUSY could be broken by nonperturbative effects. We give these comments as a justification for the study of scalar zero modes and modes of very small positive mass or modes of very small imaginary mass (where the field could be rolling). However, there could  be other reasons to study such modes.
 
As the wavelength approaches the horizon size, the naive redshift formula no longer applies and one must refine the flat space analysis of the scalar field dispersion relation \cite{Hochberg:1991jy,Hochberg:1990sr}. We will carry out a numerical analysis of the behavior of long wavelength scalar field modes and investigate the  dependence of the redshift on the scalar field mass, and its coupling to gravity.
We begin with a summary of scalar fields  coupled to gravity in an FRW universe. We then review redshift and physical wavelength formulas, after which we are in position to begin our numerical analysis. We conclude with a discussion of the implications of our results. The interpretation and application of our results requires some comments. The Universe has expanded through a number of phases during its lifetime. We think we are now transitioning from a matter-dominated phase to a vacuum-dominated phase. Earlier there was radiation domination, and before that, inflation, which occurred some time before Big Bang nucleosynthesis, but it is not clear how much before. Our task is to follow modes through these phases.

It is  unlikely that long wavelength modes can be measured directly so we are obliged to consider indirect measurements.  These involve the long-wavelength background on  which CMB or other short wavelength radiation propagates, and a proper analysis would consist of comparing observation with results predicted with and without dispersion.

Our best guess for the relevant long wavelength modes that will lead to distortion of the CMB are those modes that were produced during inflation, then pushed outside the horizon during inflation, and have recently re-entered our horizon. These are the lowest $\ell$ modes. They will have their distortion preserved due of the fact that they have spent the time from which they left the horizon until the present epoch frozen, and so unable to grow or dissipate. Being of the order of the present horizon size, they will also display the maximum distortion.
We give generic results that can be applied to any model, but since results are inflationary model dependent, a full analysis would require specifying model parameters $\xi$, $m$, etc.

\section{Scalar fields coupled to gravity in an FRW universe}

We first review the properties of a real free massive scalar fields coupled to gravity. 
The most general action (to first order in $R$) for this case is
\be\label{2-1}S=\int d^{4}x\sqrt{-g}(g^{\mu\nu}\partial_{\mu}
\phi\partial_{\nu}\phi-m^{2}\phi^{2}-\xi\phi^{2} R)\,,\ee
where $R$ is the scalar curvature and $\xi$ is a dimensionless coupling. 
We will work in FRW geometry, and use the convenient conformal parametrization 
of this family of spacetimes,
\be\label{2-2}g_{\mu\nu}=C(\eta)\mbox{diag}(1,-h_{ij}(\mathbf{x}))\,,\ee
where $C(\eta)\equiv a^{2}(t)$ is the conformal scale factor related to a conformal time via
\be\label{2-3}\eta(t)=\int^{t}\frac{c\,dt'}{a(t')}\,.\ee
The spacial part of the metric is
\be\label{2-4}h_{ij}(\mathbf{x})=diag((1-\kappa r^{2})^{-1},r^{2},r^{2}\sin^{2}\theta)\,,\ee
with $\kappa=0,1,-1$ corresponding to (flat), deSitter (positive) or Anti-deSitter 
(negative) curved spatial sections, respectively.\\

The action (\ref{2-1}) leads to the field equation
\be\label{2-5}\Box\phi+m^{2}\phi+\xi\phi R=0\,,\ee
where $\Box =\frac{1}{\sqrt{-g}}\partial_{\mu}(\sqrt{-g}g^{\mu\nu}\partial_{\nu})$.
Because of the homogeneity and isotropy of the FRW metric, the solution to the field equation 
factorizes to
\be\label{2-6}\phi_k(\eta,\mathbf{x})=C^{-1/2}(\eta)f_{k}(\eta)\mathcal{Y}_{k}(\mathbf{x})\,,\ee 
where $\mathcal{Y}_{k}$ is an eigenfunction of the spatial laplacian
\be\label{slap}\frac{1}{\sqrt{-h}}\partial_{i}\left[\sqrt{-h}h^{ij}\partial_{j}\mathcal{Y}_{k}(\mathbf{x})\right]=-\left(|\mathbf{k}|^{2}-\kappa\right)\mathcal{Y}_{k}(\mathbf{x})\,,\ee
and $k=|\mathbf{k}|$.
In the massive case, the temporal part of (\ref{2-6}) has to satisfy
\be\label{tempo}\ddot{f}_{k}+\left[k^{2}+m^{2}C(\eta)+\xi-\frac{1}{6}R(\eta)C(\eta)\right]f_{k}=0\,.\ee
One can express the scalar curvature $R$ in terms of scale factor $C$ and curvature constant $\kappa$ in the form
\be\label{rc}\frac{1}{3}RC=\frac{\ddot{C}}{C}-\frac{1}{2}\left(\frac{\dot{C}}{C}\right)^{2}+2\kappa\,.\ee
Thanks to this relation, equation (\ref{tempo}) takes the form
\be\label{tempo2}\ddot{f}_{k}+\left\{k^{2}+m^{2}C+6\left(\xi-\frac{1}{6}\right)\kappa+3\left(\xi-\frac{1}{6}\right)\left[\frac{\ddot{C}}{C}-\frac{1}{2}\left(\frac{\dot{C}}{C}\right)^{2}\right]\right\}f_{k}=0\,.\ee
The theory is conformally coupled if $\xi=\frac{1}{6}$ and $m^{2}=0$.
We will concentrate on two realistic cosmological regimes: vacuum (VDU) and matter (MDU) dominated
epochs. 
In these cases (\ref{tempo2}) reduces to
\be\label{tempo3}\ddot{f}_{k}+\left[\beta^{2}+m^{2}C-\frac{\nu^{2}-\frac{1}{4}}{\eta^{2}}\right]f_{k}=0\,,\ee
 where we have introduced the index $\nu$ defined by
\be\label{nu2}\nu^{2}(\xi,p)=\frac{1}{4}-(6\xi-1)\frac{p(2p-1)}{(p-1)^{2}}\,,\ee
(with $p=2/3$ for MDU and is also formally $2/3$ for VDU) and a conformal wave number $\beta$, corresponding to a mode $k$:
\be\label{mbeta}\beta^{2}=\left[\frac{4\pi^{2}}{\lambda_{0}^{2}}+(6\xi-1)(\Omega_{0}-1)H_{0}^{2}\right]a^{2}(t_{0})\,.\ee
Here $\lambda_{0}$ denotes the physical wavelength, corresponding to the wave number $k$, as measured today, $\Omega_{0}$ is the present ratio of matter-energy density to critical density and $H_{0}$ is the present value of the Hubble parameter.
Since all current observational evidence points toward a flat universe, we set $\kappa=0$ ($\Omega_{0}=1$) so that (\ref{mbeta}) reduces to 

\be\label{mbetashort} \beta=k=\frac{2\pi}{\lambda_{0}}a(t_{0})\,. \ee

\subsection{Massless case}
In the massless case (\ref{tempo2}) reduces to
\be\label{tempomless}\ddot{f}_{k}+\left[\beta^{2}-\frac{\nu^{2}-\frac{1}{4}}{\eta^{2}}\right]f_{k}=0\,.\ee
The solutions to this equation can be written in terms of Hankel functions $H_{\nu}^{(1)}(\beta\eta)$, which in polar form are
\be\label{polar}H_{\nu}^{(1)}(\beta\eta)=A(\beta\eta)e^{-i S(\beta\eta,\nu)}\,,\ee
with $A$ and $S$ being real valued amplitude and phase functions. These are easily expressed in terms of ordinary Bessel functions $J_{\nu}$ and Bessel functions of the second kind $Y_{\nu}$. The phase is 
\be\label{ex1}S(\beta\eta,\nu)=\arctan\frac{\cot(\pi\nu)J_{\nu}(\beta\nu)-\csc(\pi\nu)J_{-\nu}(\beta\eta)}{J_{\nu}(\beta\eta)}\,,\ee
for real $\nu(\xi,p)$ and
\be\label{ex2}S(\beta\eta,\nu)=\arctan\frac{\Im\left[e^{-i\nu\pi}J_{\nu}(\beta\eta)-J_{-\nu}(\beta\eta)\right]}{\Re\left[e^{-i\nu\pi}J_{\nu}(\beta\nu)-J_{-\nu}(\beta\nu)\right]}\,,\ee
 for imaginary $\nu(\xi,p)$, while the amplitude is 
\be\label{amp}A(\beta\eta,\nu)=\sqrt{J_{\nu}^{2}(\beta\eta)+Y^{2}_{\nu}(\beta\eta)}\,,\ee
for real $\nu$ and for imaginary $\nu$ we find
\be\label{ampim}A(\beta\eta,\nu)=|\csc(\pi\eta)|\sqrt{\left\{\Re\left[e^{-i\pi\nu}J_{\nu}(\beta\eta)-J_{-\nu}(\beta\eta)\right]\right\}^{2}+\left\{\Im\left[e^{-i\pi\nu}J_{\nu}(\beta\eta)-J_{-\nu}(\beta\eta)\right]\right\}^{2}}\,.\ee
The angular frequency of FRW modes associated with a wave number $k$ is given by
\be\label{omdef}\omega_{k}=\frac{\partial{S}}{\partial\eta}\,,\ee
where $S$ is the corresponding phase given by either (\ref{ex1}) or (\ref{ex2}), depending on the choice of cosmology.

\subsection{Massive case}
We want to write (\ref{tempo2}) in the form $\ddot{f}_{k}+\omega_{k}^{2}f_{k}=0$. Therefore using a WKB analysis one finds the frequency (\ref{omdef}) to second order $\ddot{f}_{k}+\omega_{k}^{(2)\,2}f_{k}=0$, where $\omega_{k}^{(2)}$ is given \cite{Hochberg:1991jy} by
\be\label{omega2}\omega_{k}^{(2)}(\eta)=\omega_{k}^{(0)}+\frac{3\xi-\frac{1}{2}}{2\omega_{k}^{(0)}}
\left[\frac{\ddot{C}}{C}-\frac{1}{2}\left(\frac{\dot{C}}{C}\right)^{2}\right]
-\frac{m^{2}}{8(\omega_{k}^{(0)})^{3}}\left[\ddot{C}-\dot{C}\frac{\dot{\omega}_{k}^{(0)}}{\omega_{k}^{(0)}}-
\frac{3m^{2}}{4}\frac{\dot{C}^{2}}{(\omega_{k}^{(0)})^{2}}\right]\,,\ee
with
\be\label{omega0}\omega_{k}^{(0)}=\sqrt{\beta^{2}+m^{2}C(\eta)}\,.\ee

We have checked the validity of the WKB approximation (see appendix-\ref{app:a}) by comparing with the massless limit where exact solutions are available.
The zeroth order contributions to the frequency and conformal scale factor and their derivatives for the cosmological cases of interest are summarized in Table-\ref{app-vac}\footnote{In the following we take the value of the  present energy density to be $\rho_{0}=\rho_{crit}=9.21\times 10^{-27}\frac{kg}{m^{-3}}\Rightarrow H_{0}^{-1}={a_{vac}^{0}}^{-1}=4.42\times 10^{17}s$.}.

\begin{table}[!h]
\bc
\begin{tabular}{|l|l|l|l|l|l|l|}
\hline
& $\ds{a(t)}$ & & &$\eta\left[a(t)\right]$ & $\eta(z)$\\
\hline
VDU & $\ds{e^{a_{vac}^{0}t}}$ & 
$\ds{{a_{vac}^{0}}=\left[\frac{8\pi G_{N}}{3}\rho_{0}(w)|_{w=-1}\right]^{\frac{1}{2}}}$ & 
$\ds{{a_{vac}^{0}}\simeq 2.27\times 10^{-18}s^{-1}}$& 
$\ds{-\frac{c}{{a_{vac}^{0}}a(t)}}$ &  
$\ds{-c\,{a_{vac}^{0}}^{-1}(z+1)}$ \\
\hline
MDU & $\ds{a_{mat}^{0}t^{\frac{2}{3}}}$ &
$\ds{{a_{mat}^{0}}=\left[6\pi G_{N}\rho_{0}(w)|_{w=0}\right]^{\frac{1}{3}}}$ &
$\ds{{a_{mat}^{0}}^{\frac{3}{2}}\simeq 2.05\times 10^{- 18}s^{-1}}$ &
$\ds{\frac{3c}{{a_{mat}^{0}}^{2}}a(t)}$ &
$\ds{2c\,{a_{mat}^{0}}^{-1}(1+z)^{-\frac{1}{2}}}$ \\
\hline
\end{tabular}
\ec
\caption{\label{app-vac}{Important factors for VDU and MDU.}}
\end{table}
The scale factor and hence conformal time depends on the given epoch, so for power law expansions we have
\be\label{vaot}a(t)=\left[6\pi G_{N}(1+w)^{2}\rho_{0}(w)t^{2}\right]^{\frac{1}{3(1+w)}}\ee
as can be directly determined from the Friedmann equation, where $w$ is the proportionality constant in the equation of state $P(t)=w\rho(t)$ appropriate for a given background, and $\rho_{0}(w)$ is a present value of critical density for a given epoch.
Relevant choices of parameters for use in (\ref{omega2}) and (\ref{vaot}) are given in Table-\ref{app-vac} and Table-\ref{app:1}.

\begin{table}[!h]
\bc
\begin{tabular}{|c||c|c|c|c|c|c|}
\hline
 Epoch: $(w,p)$         & $\omega_{k}^{(0)}$                & $\dot{\omega}_{k}^{(0)}$ & $a(t)$ & $C(\eta)$ & $\dot{C}(\eta)$ & $\ddot{C}(\eta)$\\
\hline\hline
Vacuum: $\ds{\left(-1,\frac{2}{3}\right)}$   & $\ds{\sqrt{\beta^{2}+m^{2}\left(\frac{c}{a_{vac}^{0}}\right)\eta^{-2}}}$ & $\ds{-\frac{m^{2}\eta^{-3}}{\omega_{k}^{(0)}}}$ & $\ds{\exp\left\{\left[\frac{8\pi
G_{N}}{3}\rho_{0}\right]^{\frac{1}{2}}t\right\}}$ & $\ds{\left(\frac{c}{a_{vac}^{0}}\right)^{2}\eta^{-2}}$   & $\ds{-2\left(\frac{c}{a_{vac}^{0}}\right)^{2}\eta^{-3}}$ & $\ds{6\left(\frac{c}{a_{vac}^{0}}\right)^{2}\eta^{-4}}$ \\ \hline
Matter: $\ds{\left(0,\frac{2}{3}\right)}$  & $\ds{\sqrt{\beta^{2}+m^{2}\frac{{a_{mat}^{0}}^{6}}{81c^{4}}\eta^{4}\eta^{4}}}$  & $\ds{2\frac{m^{2}\eta^{3}}{\omega_{k}^{(0)}}}$ & $\ds{\left[6\pi G_{N}\rho_{0}\right]^{\frac{1}{3}}t^{\frac{2}{3}}}$ & $\ds{\frac{{a_{mat}^{0}}^{6}}{81c^{4}}\eta^{4}}$ & $\ds{4\frac{{a_{mat}^{0}}^{6}}{81c^{4}}\eta^{3}}$ & $\ds{12\frac{{a_{mat}^{0}}^{6}}{81c^{4}}\eta^{2}}$ \\
\hline
\end{tabular}
\ec
\caption{\label{app:1}{Zeroth order result for frequency, scale factor and their derivatives, where a given epoch is characterized by: $w$, the proportionality constant in the equation of state $P(t)=w\rho(t)$ appropriate for a given background,
as well as the exponent of a power-law type cosmologies, i.e. $a(t)\sim t^{p}$.}}
\end{table}

\subsection{Redshift formula}
The classical redshift formula in terms of frequency $\nu$ is
\be\label{naive}\frac{\nu_{0}}{\nu}=\frac{a(t)}{a(t_{0})}\,.\ee
We want to find the correction factor to this naive redshift formula where the correction is the result of the nontrivial modifications to the dispersion relations for long-wavelength modes i.e., wavelengths of the order of the horizon size. 
To do this we have to take into account of the conformal angular frequency correction \cite{Hochberg:1990sr} so that we find 
\be\label{notnaive}\frac{\nu_{0}}{\nu}=\frac{a(t)}{a(t_{0})}\frac{\omega_{k}(t_{0})}{\omega_{k}(t)}\,.\ee
\begin{table}[!h]
\bc
\begin{tabular}{|l||c|c|}
\hline
& Conformal & Physical\\
\hline\hline
space        & $x$                                      & $x_{phy}=a(t)x$\\
\hline
time         & $\eta=\displaystyle{\int^{t}\frac{c\,dt'}{a(t')}}$        & $t$\\
\hline
wave vector  & $k$                                      & $\displaystyle{k_{phy}=\frac{k}{a(t)}}$\\
\hline
wavelength  & $\displaystyle{\lambda=\frac{2\pi}{|k|}}$               & $\displaystyle{\lambda_{phy}=a(t)\lambda}$\\
\hline
frequency    & $\displaystyle{\omega=\frac{\partial S}{\partial\eta}}$ & $\displaystyle{\omega_{phy}=\frac{\omega}{a(t)}}$\\
\hline
\end{tabular}
\ec

\caption{\label{app:2}{Dictionary of conformal and physical variables.}}
\end{table}
Here for $\omega_{k}$ we use $\omega_{k}^{(0)}=\beta$ in the massless case and $\omega_{k}^{(2)}$, given by equation (\ref{omega2}), in the massive case. In the following sections we are going to present the results for two different cosmologies. The relations between physical and conformal variables are summarized in Tables \ref{app-vac} and \ref{app:2}.
The advantage of an analysis via the WKB method is that it is simple and straightforward, and it usually gives the correct trends when the corrections are 
moderate ($\sim5\%$ to  $\sim20\%$) (as we have shown in Appendix-\ref{app:a}). These observations can be verified by comparing with exact results where they exist. In the cosmological regimes of relevance (vacuum and matter domination), we find the dependence of the dispersion relations  on the value of the mass of the scalar field. We formulate our discussion in terms of conformal invariance, i.e.,  in terms of the value of the mass and wavelength where the conformal behavior is approximately preserved.

In all cases we find  $m\sim 10^{-33}eV$ (inverse Hubble size) as the mass where nonconformal behavior starts to set in. These masses should be nearly equal in the different regimes, since differences are caused by numerical factors of order one.

To proceed further, we first have to express all the parameters present in eqs. (\ref{omega0}) and (\ref{omega2}) in terms of redshifts
\be\label{red}z=\frac{a(t_{0})}{a(t)}-1\,,\ee
 and the parameter $b_{0}$
\be\label{b0}b_{0}\equiv\frac{\lambda_{0}}{c H_{0}^{-1}}\,,\ee
that can be interpreted as the fractional size of the physical wavelength in the units of the present Hubble radius. We will work in units where $a(t)$ and hence $C(\eta)$ are dimensionless. The wave number $\beta$ does not depend on the epoch, as can be seen from equation (\ref{mbetashort}).
We have now collected all the necessary epoch specific input needed for our numerical analysis that will be carried out in the next section.

\section{Correction to Redshifts}
\begin{figure}[!h]\center
\includegraphics[width=17.5cm,height=3.8cm]{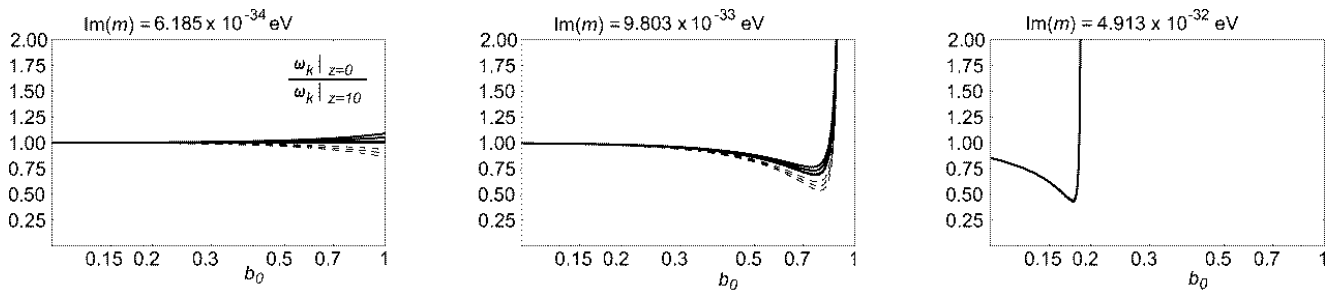}
\caption{Vacuum Domination $m^{2}<0$ WKB: Dashed curves are for $\xi<1/6$, and thick curves are for $\xi>1/6$.}\label{vac-2}
\end{figure}

\begin{figure}[!h]\centering
\includegraphics[width=17.5cm,height=3.8cm]{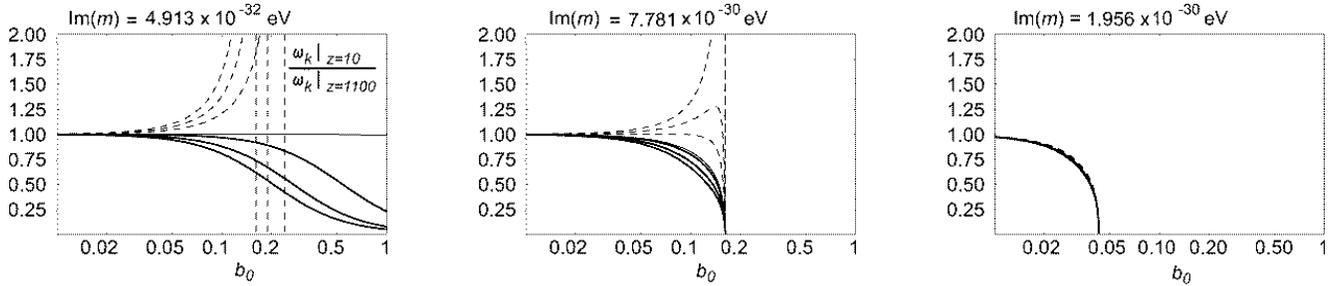}
\caption{Matter Domination $m^{2}<0$ WKB: Dashed curves are for $\xi<1/6$, and thick curves are for $\xi>1/6$.}\label{mat-2}
\end{figure}

In this section we are going to present corrections due to the redshift formula (\ref{notnaive}) originating from the coupling of the scalar fields to gravity. We consider massless as well as massive cases in two different cosmologies, i.e., vacuum and matter dominated universes.
In order to see the full spectrum of possible behavior of the dispersion relations, one has to discuss both real and imaginary masses. We plot the ratio of initial to final frequencies in each epoch. Sequentially through matter and vacuum domination, we set $z^{mat}_{i}=1100$, next $z^{mat}_{f}=z^{vac}_{i}=10$ and finally $z^{vac}_{f}=0$.
\noindent
\begin{figure}[!h]\centering
\includegraphics[width=17.5cm,height=3.8cm]{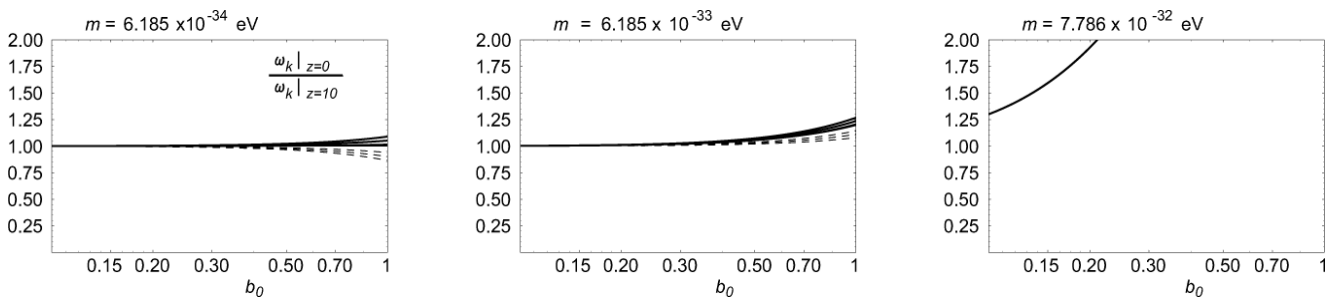}
\caption{Vacuum Domination $m^{2}>0$ WKB: Dashed curves $\xi<1/6$, Thick curves $\xi>1/6$. In this and the following figures we use $\xi=\pm 3/4,\pm 1/2,\pm 1/4,0$ for $\xi$s.}\label{vac-1}
\end{figure}
\noindent
In the case of vacuum domination, for real masses (Fig-\ref{vac-1}), the deviation from the classical redshift formula is very small until length scales of the size $\sim 0.6c\,H_{0}^{-1}$. For imaginary mass, when $\Im(m)\lesssim 10^{-32}eV$, the ratio of frequencies exhibits similar behavior (Fig-\ref{vac-2}). 
In both cases, as the magnitude of the mass of a scalar field gets larger than roughly $10^{-32}$ eV,  it dominates the effects of $\xi$ if $\xi\lesssim 1$.
\noindent
\begin{figure}[!h]\centering
\includegraphics[width=17.5cm,height=3.8cm]{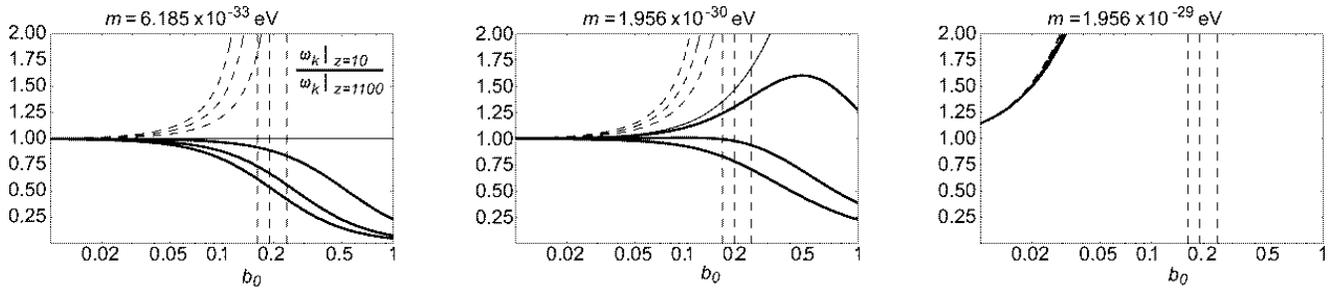}
\caption{Matter Domination $m^{2}>0$ WKB: Dashed curves are for $\xi<1/6$, and thick curves are for $\xi>1/6$. On this and the following figures vertical lines show the corresponding asymptotes where the WKB approximation fails.}\label{mat-1}
\end{figure}
\noindent
In the matter dominated case, both real (Fig-\ref{mat-1}), and imaginary (Fig-\ref{mat-2}) masses of order $\sim|10^{31}|eV$ dominate effects when $\xi\lesssim 1$. The corrections become substantial at smaller length scales $\sim 0.1c\,H_{0}^{-1}$. 
\noindent


\section{WMAP fit and Discussion}

Now let us discuss the processing of the density perturbation spectrum. As is well known, once a perturbation comes within the horizon, it begins to oscillate. This phenomenon is reflected in the observed large $\ell$ WMAP CMB spectrum, where the first maximum (first acoustic peak) is at $\ell \sim 200$, and higher order peaks are at larger $\ell$ value. The low $\ell$ values have not been inside the horizon long enough for much processing to have taken place. The region from roughly $\ell=20 $ to $\ell=200 $ is just now beginning to undergo its first plasma oscillation, while for $\ell \lesssim 20 $ very little has happened yet. But this is just the region of interest for the dispersive effects we have been discussing, and we are lucky that in this region ($\ell \lesssim 20 $) we have a pristine, unprocessed spectrum. (Recall for large $\ell$, we have short wavelengths, and so virtually no dispersion.) Hence we can confine ourselves to the analysis of perturbations with wavelengths of the order of or somewhat less than the horizon size where we do not need to worry about plasma oscillations. (There may be issues of re-ionization to consider, but at the level we are working we choose to ignore such effects.)

\begin{figure}[!h]\centering
\includegraphics[width=6.8cm]{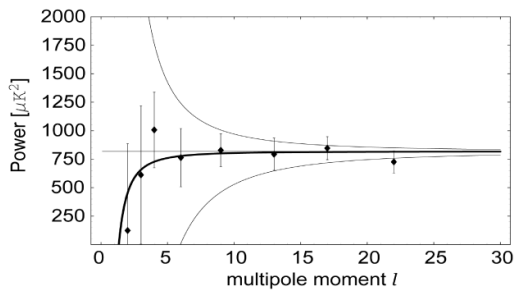}
\caption{Fit of the low $\ell$ part of WMAP spectrum to scalar fields with dispersion. The thick line is for the best fit value, $\xi = 0.166435$. We have added three other curves for comparison. The upper thin line is for $\xi = 0.169492$, the flat line is for $\xi=1/6$, and the lower thin line is for $\xi = 0.161290$.}\label{wmap-1}
\end{figure}

\begin{figure}[!h]\bc
\includegraphics[width=6.8cm]{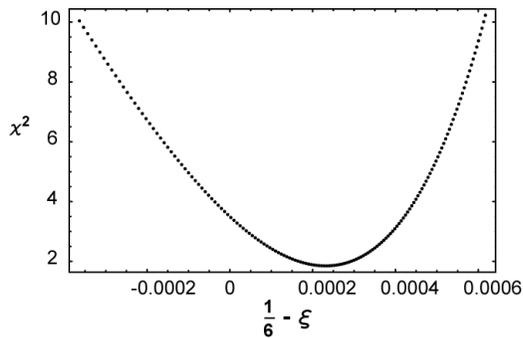}
\caption{Deviation of $\xi$ from the conformal value of coupling $\xi=\frac{1}{6}$ as seen on a plot of $\chi^{2}$ vs. $(\frac{1}{6}-\xi)$.}\label{chi2}
\ec\end{figure}

Perturbations that are of the order of the horizon size today have undergone an evolution from the time of their production. A typical scenario would be: Perturbations are produced during a vacuum-dominated epoch of inflation, and are pushed outside the horizon. Inflation then ends, and the universe becomes radiation dominated. Some perturbations come back inside the horizon, are processed via plasma oscillations, etc., until about $z=1100$, when the universe becomes matter dominated. More perturbations re-enter the horizon and are processed until around $z=10$ when the universe again becomes vacuum dominated and perturbations again start to be pushed outside the horizon. Ultimately, we would like to follow the entire evolution of a perturbation from its production until today for modes that are currently near horizon size (an $\ell \lesssim 20$ mode). 
However, this would require a detailed model of the early universe. A less ambitious approach is to follow some (better understood) fraction of the evolution to demonstrate that dispersion can play a role in understanding the observational data, and leave it to future work, when a more detailed understanding of the early Universe including details of early Universe phase transitions are known, to follow the complete evolution of the modes. To this end we have shown in Figure (\ref{wmap-1}) (where we convert wavenumber $k$ to multi-pole moment $\ell$ using $k=\ell/(c\,H_{0}^{-1})$) the results of evolving modes from\footnote{For more precise $z$ values see \cite{Hu:1995en}.} $z=1100$ until today (i.e. from $z=1100$ to $z=10$ with matter domination, and from $z=10$ until today ($z=0$) with vacuum domination), and have fit the results to the low $\ell$ WMAP data. Since an $\ell=20$ mode undergoes very little processing or dispersive evolution, we have normalized our amplitude to this region of the spectrum. Once this is done, we have a single free parameter $\xi$ for massless scalar fields $\phi$. We then do a one-parameter fit, as shown in Fig-\ref{wmap-1}, and find $\xi\simeq \frac{1}{6}-0.0002$ see Fig-\ref{chi2}. This is very close to the conformal value $\xi=\frac{1}{6}$ and can be shifted there, but only at the expense of of introducing a small negative mass squared for the scalar field as can be seen from the field equation (\ref{tempo2}) evaluated for the VDU
\be\label{2-7}\ds{\ddot{f}_{k}(\eta)+\left[\beta^{2}+\frac{(12\xi-2)+(m c/a_{vac}^{0})^{2}}{\eta^{2}}\right]f_{k}=0\,.}\ee
It is clear that one can introduce the effective coupling $\bar{\xi}$ such that
\be\label{2-8}\ds{12\bar{\xi}=12\xi+\left(\frac{m c}{{a_{vac}^{0}}}\right)^{2}}\,,\ee 
and by setting $\bar{\xi}$ to the conformal value $1/6$ we can find the value of mass $m$ corresponding to a field $\phi$ with an effective conformal coupling to gravity. 
Our conclusion is that the evolution of large $\ell$ scalar modes can be used to constrain the coupling of scalar fields, in particular the inflaton, to gravity. If we set $m=0$ we find that minimally coupled field ($\xi=0$) is easily excluded. Our fit is merely an example of how constraints on $\xi$ can arise. Specific models will give specific results. It is interesting that, with a few assumptions, a value of $\xi$ can in principle be extracted from a study of the cosmic microwave background. The coupling of scalar fields to gravity have other ramifications that would need to be considered in any realistic model. 
Another reason for being cautious about drawing sweeping inferences from Fig-\ref{wmap-1} is that the action given in (\ref{2-1}) with $m^2=0$
leads to a spectral index in disagreement with the data. Perhaps what one should conclude is that we need a theory with a sufficiently complicated potential $V(\phi)$ that density perturbations can be laid down when the $\phi$ mass is sufficiently large (see \cite{Liddle:2000cg}), but where the late time effective theory is nearly conformally invariant.
\noindent
\section*{Acknowledgement}
We thank Anjan Sen, Tom Weiler and Roman Buniy for helpful discussions. This work was supported in part by U.S. DoE grant \#~DE-FG05-85ER40226.

\appendix
\section{Validity of WKB approximation}\label{app:a}
\begin{figure}
\begin{center}
\subfigure[]{\label{vacpro}\includegraphics[height=3.8cm]{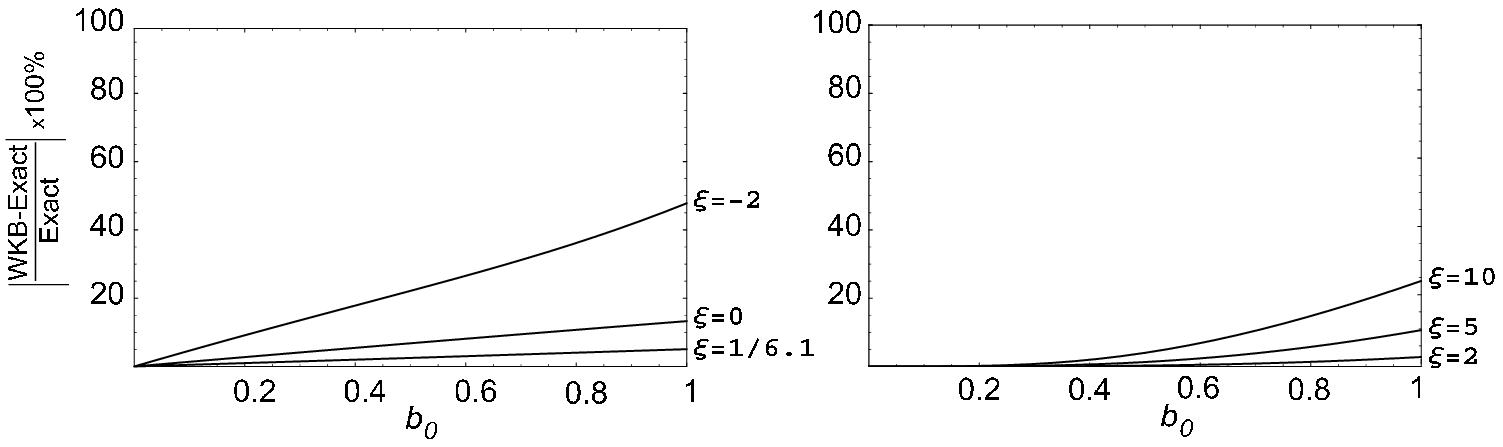}}
\hspace{1in}
\subfigure[]{\label{matpro}\includegraphics[height=3.8cm]{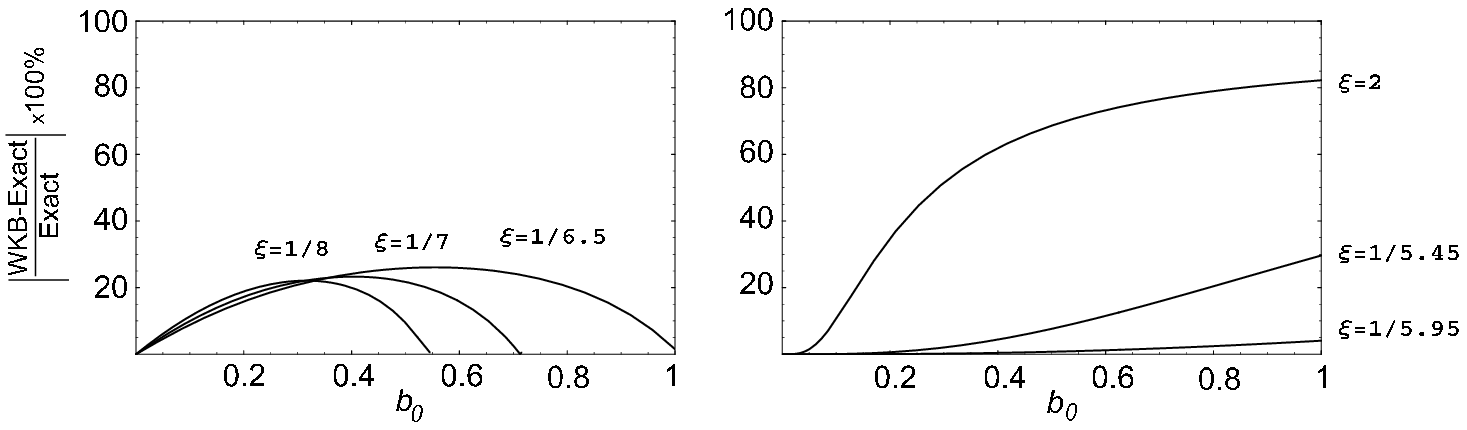}}
\end{center}
\caption{Percent error in VDU (a) and MDU (b) for both negative and positive values of $\xi$.}
\end{figure}

\begin{figure}
\begin{center}
\subfigure[]{\label{wkbvac-1}\includegraphics[height=3.8cm]{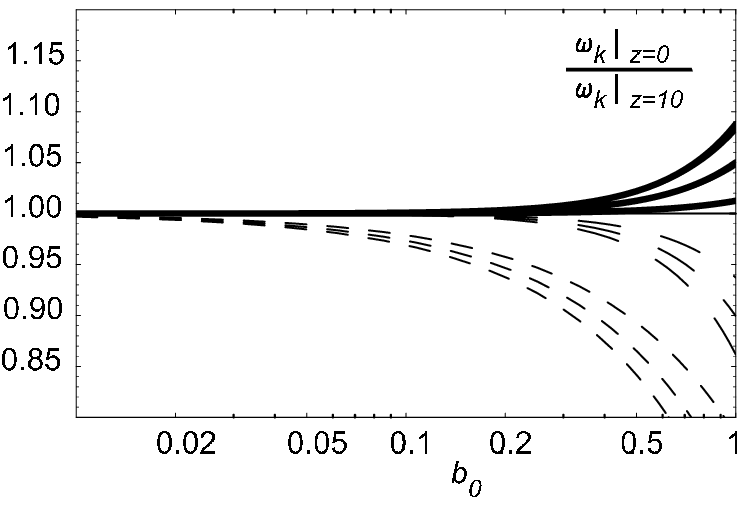}}
\hspace{1in}
\subfigure[]{\label{wkbmat-1}\includegraphics[height=3.8cm]{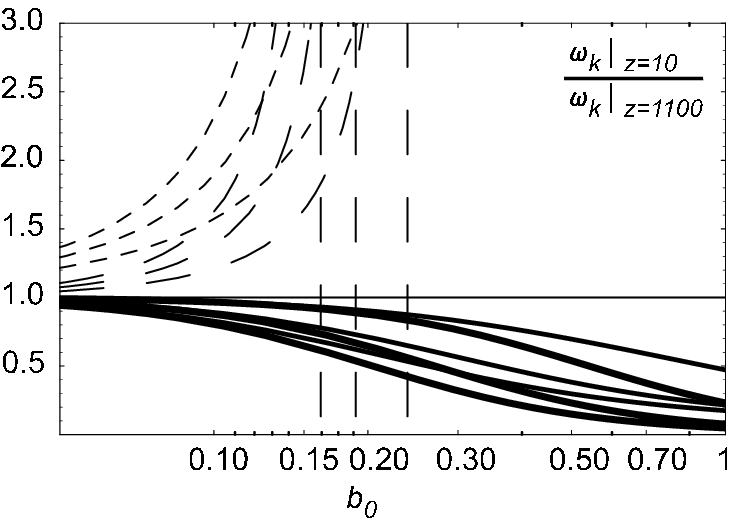}}
\end{center}
\caption{Comparison of exact solutions to the WKB approximation in VDU (a) and MDU (b) case. Thin (thick) lines represent exact (WKB) solutions with $\xi>1/6$. Small (large) dashing represents exact (WKB) with $\xi<1/6$.}
\end{figure}

The WKB approximation works best for couplings close to the conformal case $\xi=1/6$. For various choices of parameters, we show the percent errors of the WKB approximation relative to the exact vacuum dominated era results (Fig-\ref{vacpro} and Fig-\ref{wkbvac-1}), and to the matter dominated universe in (Fig-\ref{matpro} and Fig-\ref{wkbmat-1}).
\noindent
In a matter domination universe represented in Fig-\ref{wkbmat-1}, the WKB approximation is better (i.e. up to larger scales) for positive couplings. However, the broad range of applicability of the WKB method allows us to use the WKB approximation to draw conclusions about trends in the data. 

\end{document}